# Near- and middle-ultraviolet reconfigurable Raman source using a record-low UV/visible transmission loss inhibited-coupling hollow-core fiber

M. Chafer [a], J. H. Osório [b], A. Dhaybi [b], F. Ravetta [c], F. Amrani [a, b], F. Delahaye [a, b], B. Debord [a, b], C. Cailteau-Fischbach [c], G. Ancellet [c], F. Gérôme [a, b], F. Benabid [a, b, *]

[a] GLOphotonics, 123 Avenue Albert Thomas, Limoges, France
[b] GPPMM Group, XLIM Institute, CNRS UMR 7252, University of Limoges, France
[c] LATMOS/IPSL, Sorbonne University, UVSQ, CNRS, Paris, France

## ABSTRACT

We report on two types of Raman laser sources emitting in the near and middle ultraviolet spectral ranges by the use of a solarization-resilient gas-filled inhibited-coupling (IC) hollow-core photonic-crystal fiber (HCPCF) with record low transmission loss (minimum of 5 dB/km at 480 nm). The first source type emits a Raman comb generated in a hydrogen-filled HCPCF pumped by a 355 nm wavelength microchip nanosecond pulsed laser. The generated comb lines span from 270 nm to the near-infrared region with no less than 20 lines in the 270-400 nm wavelength range. The second type stands for the first dual-wavelength Raman source tuned to the ozone absorption band in the ultraviolet. Such dual-wavelength source emits at either 266 nm and 289 nm, or 266 nm and 299 nm. The relative power of the pair components is set to optimize the sensitivity of ozone detection in differential absorption lidar (DIAL). The source's physical package represents more than 10-fold size-reduction relative to current DIAL lasers, thus opening new opportunities in on-field ozone monitoring and mapping. Both Raman sources exhibit a very small footprint and are solarization-free.

## 1. Introduction

Ultraviolet (UV) spectral range extends from 300-400 nm in the near-UV, from 200-300 nm in the middle-UV, and down to 120-200 nm in the far-UV. It is established that UV laser sources are of great interest in many applications such as spectroscopy, biomedical, gas detection, and decontamination of water and food to cite a few [1]. In this application landscape, the need for small-footprint UV sources whose emission could cover the three UV bands is as pressing as pervasive, particularly in health and environment sectors. For example, in DNA sequencing, a single laser source that emits multiple and specific UV spectral lines would represent an extension in current cytometry spectral coverage to hitherto unexplored regimes in cellular analysis, and a gain in size-reduction, which would make laser integration in cytometry machines much more impactful. A second example of a timely and critical application is real-time and spatially-mapped ozone detection in the troposphere for its capability in assessing, via the ozone physio-chemical and kinetic dynamics, climate change, and air pollution [2-4].

So far, the DIAL technique has proved to be an efficient means to measure ozone concentration. However, current laser systems emitting within the ozone absorption line wavelength range between 215 nm and 300 nm are too cumbersome and costly for widespread and on-field deployment to assess the ozone dynamics over the whole globe. In fact, this shortage in compact and spectrally adjustable UV light sources is ubiquitous. Indeed, to emit directly in the UV range, excimer lasers are still widely used despite their cumbersomeness, high cost, and need for maintenance [5]. To overcome these limitations, frequency-conversion of high-power near-infrared to UV wavelengths via frequency tripling or quadrupling in borate-based crystals (e.g., LBO, BBO) has been introduced. However, this scheme offers emissions at a limited number of wavelengths. Additionally, the conversion efficiency does not exceed 50%, meaning that, after a third harmonic generation, only 25% of the pump power is converted. Finally, the crystals have a limited lifetime due to UV radiation, and the footprint remains set by a large NIR laser.

Another route that enables UV generation is the use of stimulated Raman scattering (SRS) in inhibited-coupling guiding hollow-core photonic crystal fibers (IC-HCPCF). The ability of IC-HCPCFs to confine gases in a micrometer-scale core for long and diffraction-free lengths allows exacerbating SRS conversion efficiency by a factor larger than ~$10^6$ compared to a simple capillary configuration [6]. This in turn enabled the generation of Raman combs with microchips laser pumps [7]. Another benefit of these fibers arises from the very small optical overlap between the core-guided mode with the fiber microstructured cladding. In addition to the enabling ultralow transmission loss levels, a suppressed optical overlap between the core mode with the silica cladding represents a promising means in mitigating solarization effects in UV light handling [8]. However, so far, the reported Raman spectra were limited to ~300 nm in the UV and were generated using NIR or visible pump lasers [9].

In this paper, we report on the development of two types of Raman sources using an 8-tube single-ring (SR) tubular-lattice (TL) IC-HCPCF optimized for UV guidance. The measured fiber loss was found to be as low as 10 dB/km in the UV-visible spectral region, which represents a 7.7-fold loss reduction compared to the values at this UV range [10]-[12]. The first source is based on a hydrogen-filled IC-HCPCF. By pumping it with a 355 nm microchip laser one could generate a comb-like spectrum spanning from 270 nm to 400 nm. In the second source, a 266 nm diode-pumped solid-state (DPSS) laser pumps an IC-HCPCF filled either with hydrogen, to generate a dual-wavelength emission at (266 nm, 299 nm), or with deuterium to generate (266 nm, 289 nm) wavelength pair, with a proper power ratio between the two spectral lines to maximize a DIAL sensitivity for ozone detection.

* Corresponding author. f.benabid@xlim.fr

## 2. IC-HCPCF fabrication and characterization

Fig. 1 summarizes the optical transmission properties of the IC-HCPCF we report in the present work. The fiber fabrication process was optimized for record low-loss optical guidance in the UV-visible spectral range. Fig. 1a shows the transmission loss spectrum in the 250-900 nm range. The loss spectrum has been obtained by a cutback measurement using fibers with lengths of 104.5 m and 8.5 m and a supercontinuum source (blue solid curve). To cover the full wavelength range, two optical spectrum analyzers were used. Furthermore, cutback measurements were also carried out using lasers emitting at 355 nm and 266 nm (red symbols). The inset in Fig. 1a shows a micrograph of the IC-HCPCF. The fiber was fabricated by the stack-and-draw technique. The fiber cladding is composed of 8 non-touching tubes with a diameter of 11 µm and a thickness of 600 nm. The tubes are arranged to form the surround of a hollow core with a diameter of 27 µm. Numerical simulations show that the optical overlap fraction of the fundamental core mode with the silica core contour ranges between $10^{-5}$ and $10^{-6}$ in the 300-700 nm wavelength interval [13], which is a promising feature to avoid solarization effects.

The loss measurement results show values as low as 15.7 dB/km at 725 nm and 5 dB/km at 480 nm. The latter loss figure is the lowest one ever reported for any optical fiber guiding in the blue spectral region. It is noteworthy that the measured loss values in the IC-HCPCF transmission bands between 250 nm and 600 nm are below the fundamental Rayleigh scattering limit in bulk silica (light blue-filled and dashed-contour curve). Furthermore, the loss spectrum shows two other low-loss bands in the UV range, corresponding respectively to the fiber third and fourth high order transmission bands. They are centered at 375 nm, with a minimum loss of 10 dB/km, and 245 nm with a minimum loss of 50 dB/km. The loss spectrum is consistent with the loss figures measured using 355 nm and 266 nm lasers (red symbols).

Fig. 1b summarizes the fiber UV-light handling and resilience against solarization. The blue curve shows the transmission ratio of an 8 m long IC-HCPCF when a beam from a 355 nm microchip laser (1 kHz repetition rate, 5 mW average power) has been coupled at full energy (5 µJ) to an 8 m-long fiber. The transmission was monitored over 6 months. The transmission coefficient shows a stable behavior as illustrated by a normal distribution with an average of 80% and a standard deviation of 1% (inset in the bottom right of Fig. 1b). The fiber-output beam profile was also monitored regularly in the 6 months time transmission monitoring. The results show a Gaussian-like profile consistent with the fundamental fiber core mode. To put this result into context, we observed the transmission handling of a 1 m long and 8 µm core-diameter SMF (red curve). Here, only 0.5 µJ of the 355 nm laser was coupled into the fiber. The results show that the transmission irreversibly dropped to 15% after 24h.

## 3. Demonstration of HCPCF-based UV sources

To experimentally demonstrate the UV sources referenced above, the tips of two sections from the fabricated and characterized IC-HCPCF were mounted into gas cells and loaded with a Raman gas. The formed photonic microcells are then implemented in optical setups for Raman source generation. For the Raman comb demonstration, a 355 nm microchip laser (emitting 0.8 ns pulses with a repetition rate of 1 kHz) has been used. For the ozone DIAL application, we used a laser emitting 1 ns pulses at 266 nm (with maximum energy of 40 µJ and 1 kHz repetition rate). The setup comprises standard optics for beam steering, collimation, and coupling, and a set of polarizers and waveplates for laser-beam polarization control, necessary to regulate the Raman spectral structure [14]. The coupling efficiency is typically higher than 90%.

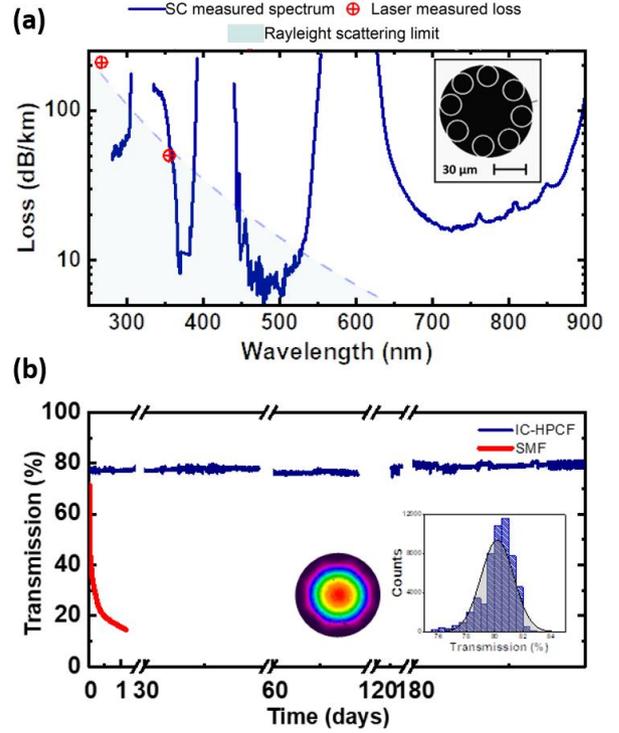

Fig. 1. (a) Measured transmission loss spectrum. The blue solid curve represents the data recorded using supercontinuum and optical spectrum analyzer. The symbols represent the data obtained using laser sources and a power-meter. (b) Long-term power stability over 6 months time. Insets show a typical near field output profile and a histogram on the fiber transmission bands.

### 3.1 UV comb pumped by a 355 nm microchip laser

For the UV comb generation, the fiber has been filled with hydrogen, and the laser energy was set to its maximum level (15 µJ). To optimize the UV components of the spectrum, a systematic study has been performed by changing the hydrogen pressure and fiber length. The spectra have been measured via free-space coupling to a photo-spectrometer with a wavelength coverage of 190 nm-1100 nm and a resolution lower than 2 nm.

Fig. 2a shows the generated stimulated Raman scattering spectra for pressure levels of 5, 7, 11, and 15 bar, respectively. Here, the fiber length was fixed to 180 cm. The laser polarization is initially set to a circular state for rotational Raman enhancement. Then, it was finely tuned to maximize the power of the spectral components in the short wavelength range so to compensate for the small polarization change during the propagation in the fiber. All the recorded spectra show discrete components corresponding to the different orders of Stokes and anti-Stokes rotational and vibrational Raman transitions. A Raman line is labeled by the integer-couple $(\pm n, \pm m)$, where $n$ and $m$ correspond to the Stokes or anti-Stokes order of the Raman vibrational and rotational transition respectively. The negative and positive signs relate to Stokes and anti-Stokes respectively. The shadowed regions in Fig. 2 identify the fiber transmission bands.

The results show that, when the pressure is increased, the conversion is shifted towards longer wavelengths. This is partly explained by the fact that the wavelength of the higher-order vibrational Stokes lines are generated in the fiber low-loss region (*e.g.*, $\lambda_{(-1,0)} = 504\ nm$, $\lambda_{(-2,0)} = 637\ nm$ and $\lambda_{(-3,0)} = 867 nm$) and the anti-Stokes ones lie in higher loss regions (*e.g.*, $\lambda_{(1,0)} = 310\ nm$ and $\lambda_{(2,0)} = 274\ nm$). Hence, to limit the gain of the Stokes transitions and, thus, to increase the number of Raman lines in the UV we need to operate pressure levels below 7 bar. Furthermore, at this pressure range, the magnitude of the Raman coefficient of the rotational transition is much closer to that of the vibrational resonance when compared to higher gas pressure.

A further pathway to optimize the Raman conversion to the UV is to conveniently adjust the fiber length. Fig. 2b shows the generated spectra for four different fiber lengths, namely 60, 120, 180, and 240 cm. Here, the pressure of hydrogen was fixed at 5 bar. The results show spectra with more than 40 Raman lines between 270 nm to 910 nm. Of particular interest is the generation of the first two vibrational anti-Stokes at the wavelengths $\lambda_{(1,0)} = 310\ nm$ and $\lambda_{(2,0)} = 274\ nm$, which was enabled by the low-loss guidance of the IC-HCPCF. Furthermore, the results show that, to obtain the highest generation in the UV, the optimum fiber length is 120 cm. For longer fiber lengths, the loss in the UV becomes a limiting factor. Conversely, if the fiber length is smaller than 120 cm, the gain is not high enough to obtain a maximum of Raman transitions in the UV.

Fig. 3a shows the optimal comb spectrum together with a picture of the projected beam obtained by using a diffraction grating (top of Fig. 3a). This procedure allowed measuring the power of each line and the corresponding spectral bandwidth to calculate the spectral densities. Fig. 3b presents the spectral density of a close-up of the spectral comb in the 265-375 nm wavelength range. The results show that, in the 345-375 nm wavelength range, 4 Raman components exhibits power spectral densities higher than 200 µW/nm. In the 305-345 nm range, 5 lines display spectral densities above 50 µW/nm. Furthermore, the 7 lines spreading from 269 to 305 nm show spectral densities above 4 µW/nm. The resulted Raman comb, therefore, covers 7 lines in the middle-UV, and 14 lines in the near-UV. The lines situated beneath 300 nm show a lower spectral density but significative enough for several applications.

### 3.2 Dual-wavelength laser source for DIAL LIDAR of $O_3$

This section reports on the second Raman source. The experimental set-up is similar to the one used for the UV-comb generator, except for the pump laser, which, for this application, emits at 266 nm. Here, we explored two configurations. The first one uses a hydrogen-filled fiber and is configured to emit at 266 nm and 299 nm. In the second configuration, the fiber is filled with deuterium to emit at the wavelength pair of 266 nm and 289 nm. The choice of the laser pump wavelength and the filling gases ($D_2$ and $H_2$) are such that the emission wavelengths of the Raman source lie within the ozone Hartley absorption band.

Fig. 4a shows the normalized ozone absorption spectrum in the spectral region around 255 nm (green dashed curve). The blue and red curves show the emitted spectrum from $H_2$- and $D_2$-based Raman sources, respectively. Both spectra show a power ratio between the two components of ~1:0.3. This power ratio is set to enhance the differential absorption signal in ozone detection DIAL and was achieved by adjusting the fiber length and the gas pressure according to a systematic study that consisted of measuring the Raman source emission spectrum evolution for different gas pressure and fiber lengths. The results of this study show that, for attaining the optimal power ratio when using $D_2$ as the filling gas, one should use a 25 cm-long fiber and a gas pressure of 3.2 bar. When using $H_2$, the optimum conditions were found to be achieved when the fiber length is 7 cm length and the gas pressure 5 bar. Finally, the measured energy of the Raman sources when the fiber is pumped by 40 µJ pulses was found to be 30 µJ for the $H_2$ configuration and 32 µJ for the $D_2$ configuration.

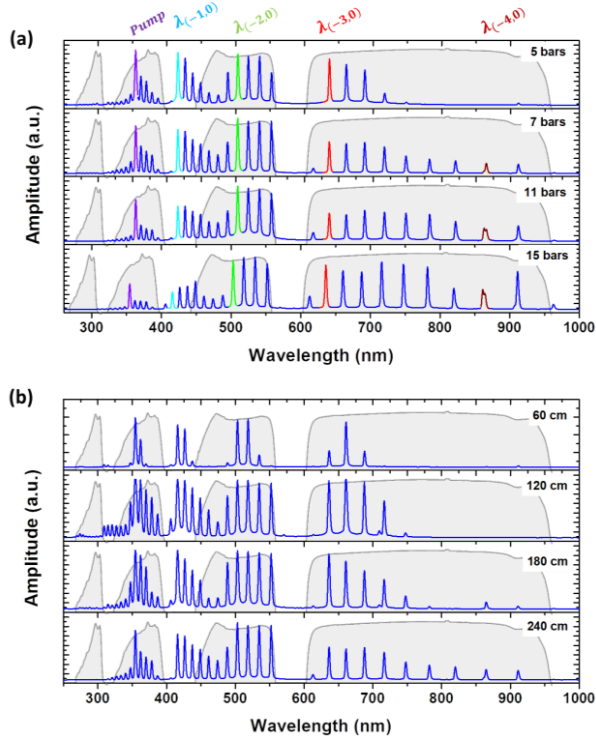

Fig. 2. (a) Comb spectra obtained by using $H_2$ at 5, 7, 11, and 15 bar for a 180 cm fiber length. (b) Comb spectra for fibers with lengths of 60, 120, 180, and 240 cm filled with $H_2$ at 5 bar.

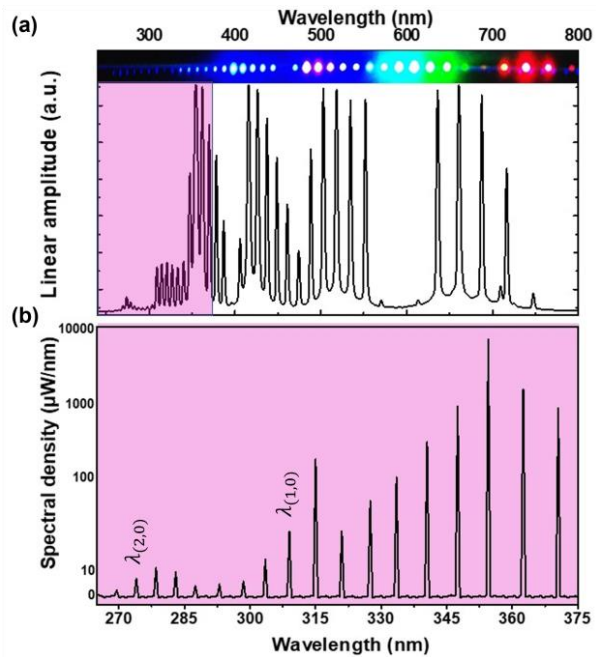

Fig. 3. (a) Comb spectrum for a fiber with length 120 cm and filled with $H_2$ at 5 bar. The diffracted output comb pattern is displayed on the figure top part. (b) Zoom in the wavelength range between 265 nm to 375 nm.

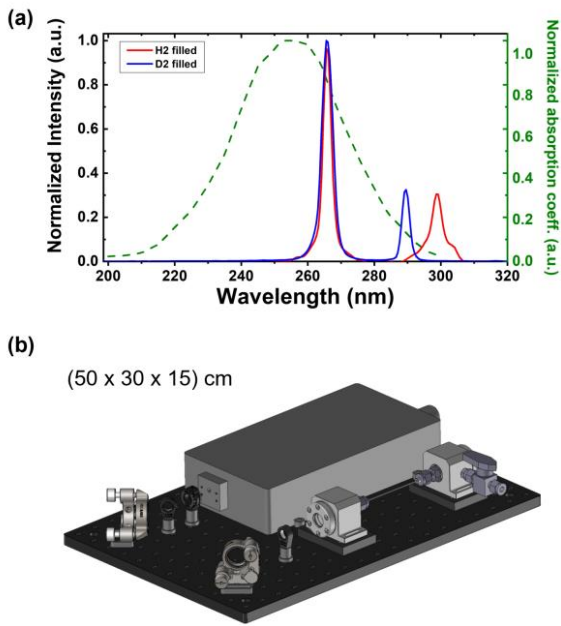

Fig. 4. (a) Raman generation plots in a 25 cm D$_2$-filled fiber (blue curve) and in a 7 cm H$_2$-filled fiber (red curve). The green curve represents the normalized absorption coefficient of ozone in the Hartley band. (b) Diagram of the HCPCF-based DIAL LIDAR source reported herein with its typical dimensions.

In both cases, the whole system sits on a breadboard of 50 cm long and 30 cm wide (Fig. 4b displays a diagram of the HCPCF-based DIAL LIDAR source reported in this manuscript). The system is, therefore, highly compact, and represents expressive size-reduction relative to current DIAL lasers, which are typically deployed in trucks due to their large sizes [15].

## 4. Conclusions

We reported on the fabrication of an IC HCPCF exhibiting record low transmission loss in the UV-visible range. The measured loss showed figures that are well below the silica Rayleigh scattering limit for all the transmission bands lying between 600 nm and 250 nm, and with a minimum value of 5 dB/km at 480 nm. The fiber was found to be very resilient to solarization and, thus, we used it to develop two Raman sources emitting in the UV range.

The robustness of the fiber to UV transmission combined with a simple set-up has led to the creation of a UV comb generator, coined the UV-Comblas, with no less than 20 lines in the UV range associated with high spectral density, which has the potential to address biomedical applications needs. Remarkably, the use of the record low-loss IC HCPCF reported herein allowed to extend the Raman comb to wavelengths as low as 270 nm while using a modest DPSS pump.

By changing the pump laser to another one emitting at 266 nm, we demonstrated the realization of a dual-wavelength laser source emitting at 266 nm and 289 nm or 266 nm and 299 nm with a relative power ratio of 1:0.3 between the 266 nm and 289 nm (or 289 nm) emissions respectively (according to the Raman active gas that fills the fiber). This source, which is the first dual wavelength Raman source tuned to the O$_3$ absorption band in the UV, is highly compact and allows to significantly reduce the footprint of the current systems, hence increasing its applicability in practical ozone concentration measurements and photochemical studies.

## Acknowledgments

This research was funded by PIA program (grant 4F).

## REFERENCES

[1] M. R. E. Lamont, Y. Okawachi, A. L. Gaeta, "Route to stabilized ultrabroadband microresonator-based frequency combs," Opt. Lett. 38, 3478 (2013).

[2] P. A. Monks, A. T. Archibald, A. Colette, O. Cooper, M. Coyle, R. Derwent, D. Fowler, C. Granier, K. S. Law, G. E. Mills, D. S. Stevenson, O. Tarasova, V. Thouret, E. von Schneidemesser, R. Sommariva, O. Wild, M. L. Williams, "Tropospheric ozone and its precursors from the urban to the global scale from air quality to short-lived climate forcer", Atmos. Chem. Phys., 15, 8889–8973, (2015).

[3] F. Ravetta, G. Ancellet, A. Colette, H. Schlager, "Long-range transport and tropospheric ozone variability in the western Mediterranean region during the Intercontinental Transport of Ozone and Precursors (ITOP-2004) campaign", J. Geophys. Res. Atmos., 112 (D10), pp.D10S46, (2007).

[4] A. Papayannis, G. Ancellet, J. Pelon, G. Mégie, "Multiwavelength lidar for ozone measurements in the troposphere and the lower stratosphere," Appl. Opt. 29, 467-476 (1990).

[5] D. Basting, K. Pippert, and U. Stamm, History and future prospects of excimer laser technology, Riken review no 43, (2002).

[6] F. Benabid, J. C. Knight, G. Antonopoulos, P. S. J. Russell, "Stimulated Raman scattering in hydrogen-filled hollow-core photonic crystal fiber," Science 298, 5592, 399-402 (2002)

[7] W. W. Duley, UV lasers effects and applications in materials science, Cambridge university press, (2005).

[8] F. Yu, M. Cann, A. Brunton, W. Wadsworth and J. Knight, "Single-mode solarization-free hollow-core fiber for ultraviolet pulse delivery," Optics Express 26, 10879-10887 (2018).

[9] A. Benoît, B. Beaudou, M. Alharbi, B. Debord, F. Gérôme, F. Salin, F. Benabid, "Over five octaves wide Raman combs in high power picosecond-laser pumped H2 filled in inhibited coupling Kagome fiber," Opt. Express 23, 11, 14002 (2015).

[10] M. Chafer, J. H. Osório, F. Amrani, F. Delahaye, M. Maurel, B. Debord, F. Gérôme, F. Benabid, "1-km hollow-core fiber with loss at the silica Rayleigh limit in the green spectral region," IEEE Photon. Technol. Lett. 31, 9, 685-689 (2019).

[11] S. Gao, Y. Wang, W. Ding, P. Wang, "Hollow-core negative-curvature fiber for UV guidance," Opt. Lett. 43, 1347-1350 (2018).

[12] F. Yu, M. Cann, A. Brunton, W. Wadsworth, J. Knight, "Single-mode solarization-free hollow-core fiber for ultraviolet pulse delivery," Opt. Express 26, 10879-10887(2018).

[13] B. Debord, A. Amsanpally, M. Chafer, A. Baz, M. Maurel, J. M. Blondy, E. Hugonnot, F. Scol, L. Vincetti, F. Gérôme, F. Benabid, "Ultralow transmission loss in inhibited-coupling guiding hollow fibers," Optica 4, 209-217 (2017).

[14] R. W. Minck, E. E. Hagenlocker, and W. G. Rado, "Stimulated pure rotational raman scattering in deuterium," Phys. Rev. Lett. 17, 5, 229–231 (1966).

[15] A. O. Langford, R. J. Alvarez II, G. Kirgis, C. J. Senff, D. Caputi, S. A. Conley, I. A. Faloona, L. T. Iraci, J. E. Marrero, M. E. McNamara, J. Ryoo, E. L. Yates, "Intercomparison of lidar, aircraft, and surface ozone measurements in the San Joaquin Valley during the California Vaseline Ozone Transport Study (CABOTS)," Atmos. Meas. Tech. 12, 1889-1904 (2019).